\documentclass[journal=jctcce,manuscript=article,layout=twocolumn]{achemso}

\usepackage{makeidx}
\usepackage{chemformula} 
\usepackage[T1]{fontenc} 

\usepackage{xr}

\usepackage{amsmath}
\usepackage{graphicx}
\usepackage{float,physics}
\usepackage{siunitx}
\usepackage{xcolor}
\usepackage{epstopdf}
\definecolor{darkgreen}{rgb}{0,0.6,0.0}

\usepackage{tabularray}
\usepackage{booktabs}
\usepackage{latexsym}
\usepackage{tabularx}
\usepackage{amsmath}
\usepackage{amsfonts}
\usepackage{amssymb}
\usepackage{mathrsfs}
\usepackage{verbatim}
\usepackage{anysize}
\usepackage{soul}
\usepackage{lipsum}
\usepackage[utf8]{inputenc}
\usepackage{algorithmic}
\usepackage{algorithm}
\usepackage{enumerate}
\usepackage{amssymb, amsmath, amsthm}
\usepackage{graphicx}
\usepackage{lmodern,url}
\usepackage{graphicx}
\usepackage{wrapfig}
\usepackage{hyperref}
\AtBeginDocument{\RenewCommandCopy\qty\SI}
\floatname{algorithm}{Algoritmo}

\newcommand{\bo}{\raise-1mm\hbox{\Large$\Box$}}

\usepackage{multirow}

\hypersetup{
    colorlinks=true,
    linkcolor=blue,    
    citecolor=blue,   
    urlcolor=blue       
}

\usepackage{xcolor}

\title[Efficient and Accurate Machine Learning Interatomic Potential for Graphene]{Efficient and Accurate Machine Learning Interatomic Potential for Graphene: Capturing Stress-Strain and Vibrational Properties}   
\author{Felipe Hawthorne}
\email{felipehawthorne@ufpr.br}
\affiliation[Federal University of Parana]{Department of Physics,
               R. Evaristo F. Ferreira da Costa, 81530-015, Curitiba, Brazil}
\alsoaffiliation[Federal University of Parana]{Interdisciplinary Center for Science, Technology, and Innovation (CICTI), Av. Cel. Francisco H. dos Santos, 81530-000, Curitiba, Brazil}

\author{Paulo R. E. Raulino}
\affiliation[Federal University of Parana]{Department of Physics,
               R. Evaristo F. Ferreira da Costa, 81530-015, Curitiba, Brazil}
\alsoaffiliation[Federal University of Parana]{Interdisciplinary Center for Science, Technology, and Innovation (CICTI), Av. Cel. Francisco H. dos Santos, 81530-000, Curitiba, Brazil}

\author{Ronaldo Rodrigues Pelá}
\affiliation{Supercomputing Department, Zuse Institute Berlin (ZIB), Takustraße 7, Berlin, 14195, Germany}

\author{Cristiano F. Woellner}
\email{woellner@ufpr.br}
\affiliation[Federal University of Parana]{Department of Physics,
               R. Evaristo F. Ferreira da Costa, 81530-015, Curitiba, Brazil}
\alsoaffiliation[Federal University of Parana]{Interdisciplinary Center for Science, Technology, and Innovation (CICTI), Av. Cel. Francisco H. dos Santos, 81530-000, Curitiba, Brazil}

\let\oldmaketitle\maketitle
\let\maketitle\relax

\begin{document}

\twocolumn[
  \begin{@twocolumnfalse}
     \oldmaketitle
  \begin{abstract}
       \textbf{Abstract}. Machine learning interatomic potentials (MLIPs) offer an efficient and accurate framework for large-scale molecular dynamics (MD) simulations, effectively bridging the gap between classical force fields and \textit{ab initio} methods. In this work, we present a reactive MLIP for graphene, trained on an extensive dataset generated via \textit{ab initio} molecular dynamics (AIMD) simulations. The model accurately reproduces key mechanical and vibrational properties, including stress-strain behavior, elastic constants, phonon dispersion, and vibrational density of states. Notably, it captures temperature-dependent fracture mechanisms and the emergence of linear acetylenic carbon chains upon tearing. The phonon analysis also reveals the expected quadratic ZA mode and excellent agreement with experimental and DFT benchmarks. Our MLIP scales linearly with system size, enabling simulations of large graphene sheets with \textit{ab initio}-level precision. This work delivers a robust and transferable MLIP, alongside an accessible training workflow that can be extended to other materials.
       \end{abstract}
  \end{@twocolumnfalse}
]

\section{Introduction}

The trade-off between accuracy and computational efficiency is a long-standing challenge in materials modeling, as summarized by the ``No Free Lunch'' theorem~\cite{Wolpert1997}. While \textit{ab initio} Molecular Dynamics (AIMD) methods based on Density Functional Theory (DFT) can achieve high precision, their computational cost with typical ${\cal O} (N^3)$ scaling, where $N$ represents the number of atoms, hinders their application to sytems containing more than tens of thousands of atoms even on high-performance computers ~\cite{Butler2016,Mueller2020}. In contrast, classical Molecular Dynamics (MD) can simulate significantly larger systems, with hundreds of thousands of atoms; however, this comes at the cost of accuracy, which heavily depends on the force field parametrization~\cite{millions,Iftimie2005}.  Force fields offer a spectrum of complexity, ranging from the computationally demanding reactive potentials~\cite{Senftle2016}, capable of reproducing bond breaking and formation, to cheaper classical potentials, without reactivity considerations~\cite{Rappe1992}. Furthermore, a non-negligible challenge in classical MD simulations is the task-specific calibration of force fields. This calibration can complicate the achievement of optimal agreement with experiment or DFT for various properties, including thermal~\cite{Lima2025}, vibrational, or mechanical~\cite{TSAI2010194}, when relying on a single parametrization optimized for a specific application.

Machine learning interatomic potentials (MLIPs) have emerged as a promising solution to bridge this gap, offering both accuracy and scalability by replacing conventional MD force fields with models trained on \textit{ab initio} datasets~\cite{Friederich2021}, enabling large-scale simulations that retain quantum-mechanical accuracy~\cite{Fedik2022, Wang2024, Unke2021}. This approach applied to materials with well-established properties and current relevancy, such as graphene~\cite{Urade2022},  offers a compelling argument for the adoption of MLIPs, while also providing a more accurate and efficient pathway to improved MD simulations. In this context, previous studies have primarily focused on investigating lattice dynamics and phonon dispersion~\cite{Rowe2018}, or on temperature-dependent mechanical properties~\cite{Singh2023}. However, to the best of our knowledge, no comprehensive effort has yet been made to develop a MLIP with DFT-accuracy that simultaneously captures both mechanical and vibrational properties, where collective and individual atomic effects are well represented, such as strain response, phonon dispersion, and shifts in vibrational spectra.

In this manuscript, we develop a MLIP for graphene and employ it to compute key mechanical and vibrational properties, such as: the elastic tensor, Young's modulus, Poisson's ratio, shear modulus, phonon dispersion curves, and vibrational density of states. The paper is organized as follows: Section \ref{sec:methods} outlines the computational methods; Section~\ref{sec:results} presents and discusses the results; and Section~\ref{sec:conclusion} summarizes the study with the conclusions.

\section{Methods}\label{sec:methods}

\begin{figure}[htb!]
   \centering
  \includegraphics[width=0.44\textwidth]{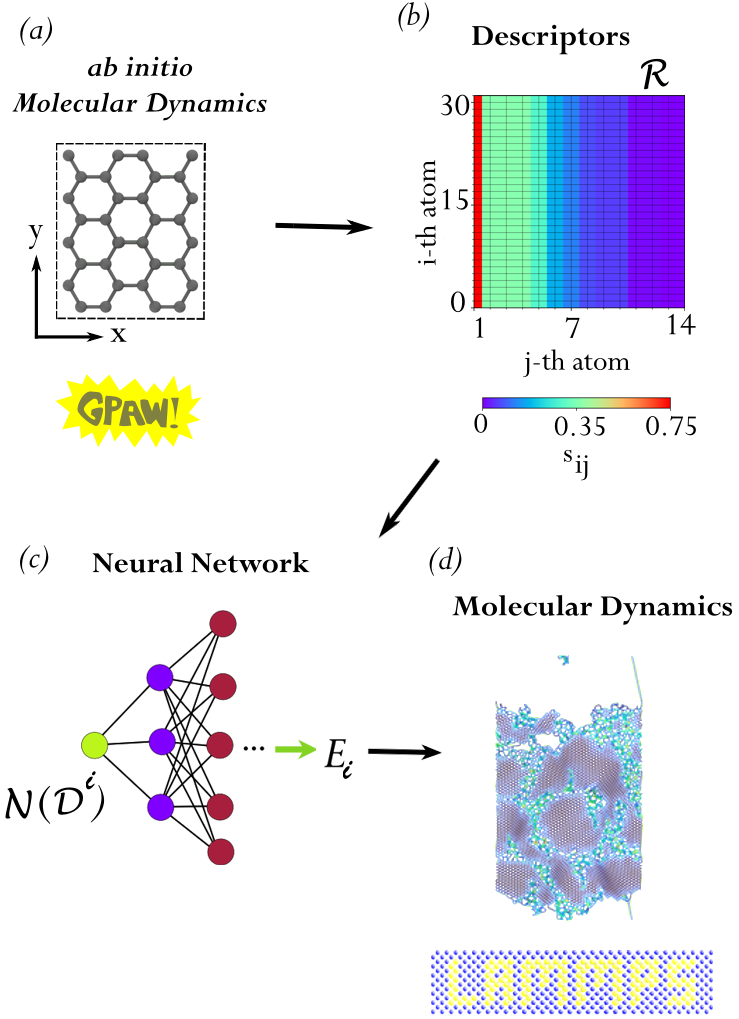}
 \caption{(a) Illustration of the 32-C graphene structure used in the AIMD simulations. (b) Generation of the coordinate matrix, ${\cal R}$, and subsequent calculation of the descriptors, ${\cal D}^i$. The first element of ${\cal R}^i$, $s_{ij}$, is illustrated. (c) These descriptors served as input elements for the NN, trained to predict the total energy $E = \sum_{i} E_i$. (d) The trained MLIP was deployed as a force field in large-scale MD simulations. The panel depicts the application of strain on a large graphene sheet at $T = 1~$K.}
\label{fig:illustration_workflow}
\end{figure}

The workflow employed for the development of the graphene MLIP, as illustrated in Figure~\ref{fig:illustration_workflow}, comprised three stages. 

As illustrated in Figure~\ref{fig:illustration_workflow}a, the relevant phase space was first explored by performing AIMD simulations on systems containing $4$, $8$, $16$, and $32$ C atoms, with the latter two considered in both $L_y > L_x$ and $L_y < L_x$ configurations, where $L_x$ and $L_y$ denote the cell's dimensions along the $x$ and $y$ directions, respectively. Figure S1a-f depicts each of these structures. These simulations included NVT, NPT, and NVE ensembles, with varying strain rates and small random atomic displacements introduced to enhance configurational diversity. In the second stage (Figure~\ref{fig:illustration_workflow}b,c), we used the DeepMD framework \cite{Zeng2023} to develop a Deep Potential~\cite{Zhang2018} based force field, using embedded atomic descriptors. Third, once satisfactory precision was achieved, the trained MLIP was deployed in large-scale MD simulations using LAMMPS~\cite{LAMMPS}, targeting stress-driven behavior. This final stage is depicted in Figure~\ref{fig:illustration_workflow}d, where the deformation of a large graphene sheet is shown under applied strain.

\textbf{\textit{Ab initio} Molecular Dynamics}. To generate a comprehensive dataset that captures the relevant potential energy surface (PES), we performed Born-Oppenheimer AIMD simulations using the projector-augmented wave method~\cite{Bloechl1994}, as implemented in the GPAW DFT calculator~\cite{Enkovaara2010, Mortensen2024} within the Python ASE framework~\cite{HjorthLarsen2017}. The simulations were conducted using a k-grid with a resolution of at least $0.05~\text{\AA}^{-2}$, a plane-wave cutoff of $500$ eV, and the Local Density Approximation (LDA) exchange-correlation functional, chosen due to its relatively lower computational cost~\cite{Ziesche1998} and its reasonable agreement with more advanced functionals in graphene-based systems~\cite{Wang2012}. AIMD simulations were performed within three thermodynamic ensembles~\cite{Frenkel2011}: $(i)$ the microcanonical (NVE) ensemble, employing the Velocity Verlet algorithm for time propagation; $(ii)$ the canonical (NVT) ensemble at $T=300~$K, using a Langevin thermostat with a friction coefficient of $10^{-2}\text{fs}^{-1}$; and $(iii)$ the isothermal-isobaric (NPT) ensemble, which allows control over temperature and pressure, enabling the application of stress along various directions and at different strain rates~\cite{Melchionna1993, Melchionna2000}.


All simulations used a timestep of $\mathrm{d}t = 0.5~\text{fs}$, and a total of $1.12\times10^6$ frames  (positions, stress, forces, energy, and cell lengths) were collected for all system sizes. Additional frames were obtained via the application of random displacements of amplitude $10^{-2} \text{\AA}$ on the atomic positions of each geometry, except for the $ 32$-atom system, where this was deemed unnecessary due to its inherent structural richness in comparison with the smaller systems. Simulation times ranged from $0.5$~ps to $5$~ps, and NPT runs contained strain rates between $10^{-6}\text{fs}^{-1}$ to $10^{-3}\text{fs}^{-1}$. During these latter runs,  tearing of the graphene structures was observed. Further details about each dataset are provided in Table S1.


\textbf{Machine Learning Interatomic Potential}. We employed the local frame two-body embedding descriptor~\cite{NEURIPS2018_e2ad76f2} within the Deep Potential framework~\cite{Zhang2018}. The local environment of each $i$-th $-$ comprising its position, the positions of its $N_{c}$ neighboring atoms, and their relative angles $-$ is defined based on a cutoff radius, $r_c$. This environment is then used to construct the coordinate matrix, ${\cal R}^i$, 
\begin{equation}
    {\cal R}^i = s_{ij} r^{-1}_{ij}\left(~r_{ij}~~x_{ij}~~y_{ij}~~z_{ij}~\right),
\end{equation}
where $s_{ij}$ is a switching function ensuring a smooth decay starting from the switching radius, $r_s$, 
\begin{equation}
    s_{ij} = \begin{cases}
         \frac{1}{r_{ij}},  \quad &r_{ij} < r_s, \\
         \frac{\left[\lambda^3\left(-6\lambda^2 +15\lambda - 10\right) +1\right]}{r_{ij}}, \quad &r_s \leq r_{ij} < r_c,\\
         0, \quad &r_{ij} \geq  r_c,\\
    \end{cases}
\end{equation}
with $\lambda =\left(r - r_s\right)\left(r_c-r_s\right)^{-1}$. 

The coordinate matrix is then used to generate a set of descriptors, ${\cal D}^i$, for each atom via auxiliary embedding residual neural networks (ResNet) \cite{Leibe2016,Chollet2021}. Each neighboring atom $j$ is processed through an embedding function $\left({\cal G}^i\right)_j$, which receives the associate $\left({\cal R}^i\right)_j$ and is trained to enrich the descriptor by vectorizing the discrete input data. The resulting descriptors can be written as
\begin{equation}
    {\cal D}^i = \left({\cal G}^i\right)^T {\cal R}^i \left({\cal R}^i \right)^T {\cal G}^i.
\end{equation}

These descriptors serve as inputs to the Deep Potential Neural Network (NN), ${\cal N}$, which is trained to predict individual atomic energies, $E_i$. The NN is optimized using AIMD data, including atomic positions, forces, the virial tensor, the simulation cell, and total energy of the system. The overall objective of the NN is to predict the total energy of the entire system, $E$,
\begin{equation}
    \sum_{i=0}^{N-1} {\cal N} \left({\cal D}^i\right) = \sum_{i =0}^{N-1}E_i = E.
\end{equation}

Once $E$ is defined as a function of ${\cal N}$, the atomic forces, $F_{i,\alpha}$, and virial tensor, $\Xi_{\alpha,\beta}$, can be obtained via differentiation as $F_{i,\alpha} = - \partial E/\partial r_{i,\alpha}$ and $\Xi_{\alpha, \beta} = - \sum_{\gamma}\left(\partial E/\partial \epsilon_{\gamma \alpha} \right)\epsilon_{\gamma\beta}$, where $i = 0, N-1$ represents the atomic index, $\alpha, \beta$ and $\gamma$ refer to Cartesian coordinates, and $\epsilon_{\alpha,\beta}$ denotes the corresponding virial direction. 

Given that part of our dataset includes strained graphene structures, we set the search and smooth decay radii to $r_c = 8~$\AA~ and $r_s = 0.5~$\AA. The size of the embedding matrix was chosen as $16$, with three hidden layers containing $50$, $100$, and $200$ neurons. The Deep Potential network consisted of three hidden layers of $240$ neurons each. The starting and final loss pre-factors~\cite{Zeng2023},  for energy, force, and virial were set as $\left(10^{-2},1,10^3\right)$,$\left(1,10^4,10\right)$, respectively. Training was conducted for a total of $10^8$ steps, using an initial learning rate of $10^{-4}$ with a decay of $5\times10^{-6}$ every $5\times10^4$ steps. $20\%$ of the data was reserved for validation.

The model's performance was assessed using the root mean square error (RMSE) computed across all validation datasets, defined as 
\begin{equation}
    \text{RMSE} = \sqrt{\langle \left( \Theta_k[{\cal N}] - \Theta_k [ \text{AIMD}] \right)^2 \rangle},
\end{equation}
where $\Theta_k$ denotes one of the physical properties (energy, force, or virial) predicted by the NN or obtained from AIMD data, evaluated over the same set of descriptors.

\textbf{Molecular Dynamics.} Classical MD simulations were performed using the LAMMPS software~\cite{LAMMPS}, to study the stress-strain behavior and the vibrational properties. Unless stated otherwise, simulations employed a rectangular graphene sheet containing 9072 carbon atoms, with $L_x = 150.77 \text{\AA}, L_y = 152.33 \text{\AA}$, depicted in Figure S1g, with a timestep of $\mathrm{d}t = 0.5~$fs.
For the stress-strain analysis, we employed two approaches. First, to obtain deformation curves up to the breaking point, uniaxial strain was applied along both the $\vec{x}$ and $\vec{y}$ directions at a strain rate of $10^{-4}\text{fs}^{-1}$, maintaining the system in an NPT ensemble at $T=1~$K  and $T=300~$K. Second, to extract the elastic tensor, an initial energy minimization at $T=0~$K ensured a mechanically stable configuration, and then a series of small finite deformations ($\epsilon_{kl}<1\%$) was applied, keeping the system within the linear elastic regime. The elastic constants were calculated using the finite deformation method, with the symmetric pressure tensor $P_{ij}$ computed as in Refs.~\cite{Surblys2019,Surblys2021}, 
\begin{equation}
    P_{ij} = \frac{1}{V}\sum_{k}^N r_{ki}f_{kj},
\end{equation}
where $V$ stands for the cell volume, $N$, the number of atoms, $r_{ki}$, the $i$-th component of the position vector of atom $k$, and $f_{kj}$, the $j$-th component of the force acting on the same atom. 
Given the definition of the stress tensor as $\sigma_{ij} = - P_{ij}$, the elastic constants were extracted using the generalized Hooke’s law, $P_{ij} = - C_{ijkl}\epsilon_{kl}$, where $ C_{ijkl}$ represents the stiffness tensor components and $\epsilon_{kl}$ the applied strain. Then, the independent elastic constants $C_{11}$ and $C_{21}$ were obtained in Voigt notation. Based on these, the Young's modulus $Y_m$ and Poisson's ratio $\nu$ were determined as $Y_m = (C_{11}C_{22}-C_{12}C_{21})/C_{22}$ and $\nu = C_{12}/C_{22}$ respectively \cite{TSAI2010194}.

In addition to mechanical properties, vibrational properties were analyzed through phonon dispersion and the vibrational density of states (VDOS). The phonon dispersion curve was obtained using Phonopy’s \cite{phonopy-phono3py-JPCM} interface with LAMMPS, using a $10\times10\times 1$ supercell based on graphene's hexagonal unit cell. The VDOS was extracted from the Fourier transform of the velocity-velocity autocorrelation function,
\begin{equation}
    C_v(t) = \left\langle \frac{\vec{v}_i(t)\cdot\vec{v}_i(0)}{\vec{v}_i(0)\cdot\vec{v}_i(0)} \right\rangle,
\end{equation}
which provides insight into the first-order Raman spectrum even without explicit projection of DFT-derived vibrational modes \cite{DICKEY1969, Thomas2013, Bosse1978}.  The VDOS calculations were performed in the NVT ensemble at $T=300~$K, with a $25$ ps thermalization step followed by $1$ ps time averages. Simulations were carried out for both the undeformed and strained graphene sheets at $\epsilon_{ii} = 1\%, 2\%, 3\%$, $i=x,y$, enabling comparison of the MLIP’s predictions with experimental observations of the red-shift in graphene’s optical spectrum upon uniaxial strain \cite{Ni2009, Dong2017, Das2008}. Fourier transforms were computed using NumPy’s \cite{harris2020array} discrete Fourier transform package.
\section{Results}
\label{sec:results}

\begin{figure}
\centering
    \includegraphics[width=0.45\textwidth]{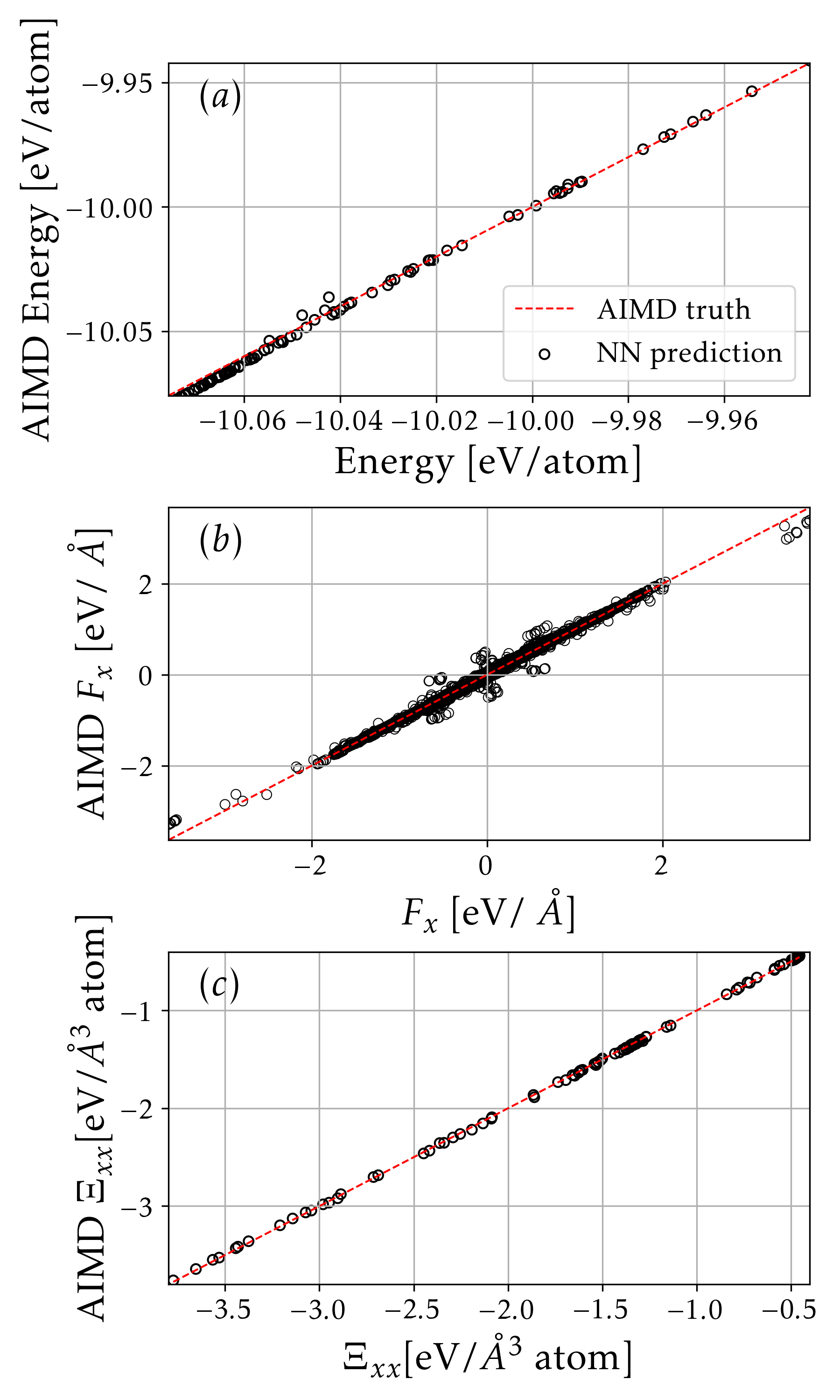}
    \caption{Comparison between AIMD reference data and NN predictions (black scattered points), derived using AIMD simulation frames from the 32-C system as input: (a) energy per atom, (b) $x$-component of atomic forces, and (c) $xx$-component of the virial stress tensor. Red dashed lines indicate perfect agreement. }
\label{fig:comparison}
\end{figure}

\textbf{MLIP training}. A direct comparison between the NN predictions trained on the 32-C validation dataset and the reference AIMD results obtained via GPAW is shown in Figure~\ref{fig:comparison}. The predicted energy per atom, force components, and virial stress exhibit only minor deviations from the AIMD reference. These deviations remain consistently low across all properties, demonstrating the NN’s ability to reproduce the interatomic interactions of graphene with high accuracy. The quantitative agreement is further supported by the RMSE values reported in Table~\ref{table:benchmark_sets}, which summarize the model’s performance across different system sizes.

\begin{table*}[h]
\centering
\caption{Root mean square error (RMSE) of the MLIP model predictions for different system sizes used in the AIMD training set. The errors are reported for energy per atom ($E_{\text{RMSE}}$), force components ($F_{\text{RMSE}}$), and the virial stress tensor ($\Xi_{\text{RMSE}}$).}
\begin{tabular*}{\textwidth}{@{\extracolsep{\fill}}ccccc}
\toprule
\toprule
$\#C$ & $E_{\text{RMSE}}$ [eV/atom] & $F_{\text{RMSE}}$ [eV/\AA] & $\Xi_{\text{RMSE}}$ [eV/\AA${}^3$ atom] \\
\midrule
$4$  & $1.94\times 10^{-2}$  & $7.74\times10^{-2}$  & $1.48\times 10^{-1}$ \\
$8$  & $2.25\times 10^{-2}$  & $5.33\times10^{-2}$  & $2.05\times 10^{-1}$ \\
$16$ & $1.44\times 10^{-2}$  & $3.68\times10^{-3}$  & $3.37\times 10^{-1}$ \\
$32$ & $8.77\times 10^{-2}$  & $5.01\times10^{-2}$  & $6.85\times 10^{-1}$ \\
\bottomrule
\bottomrule
\end{tabular*}
\label{table:benchmark_sets}
\end{table*}

\begin{figure}[h]
    \centering
    \includegraphics[width=0.45\textwidth]{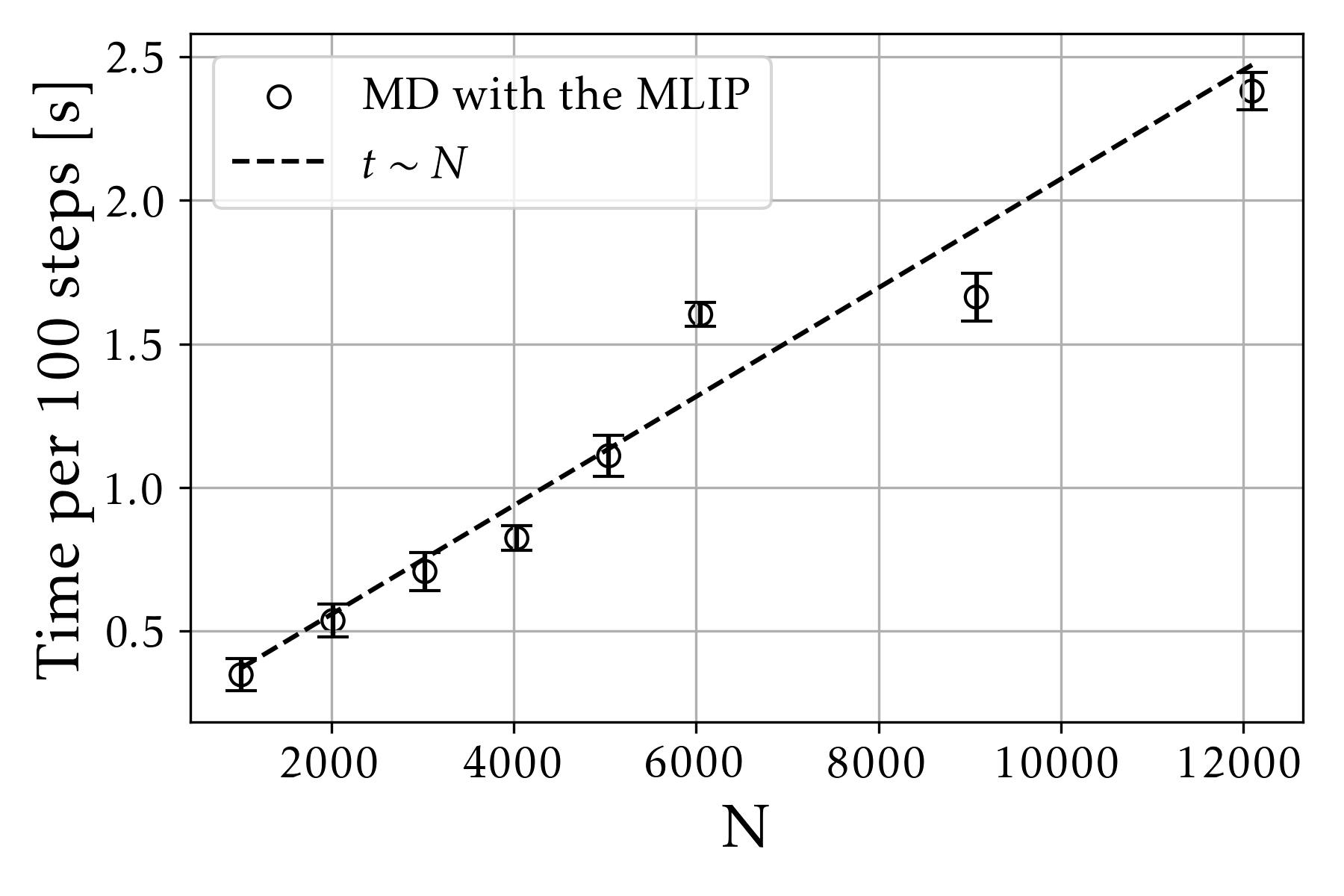}
    \caption{Average time required for $100$ NPT steps as a function of system size. The trained MLIP exhibits linear scaling behavior, ${\cal O} (N)$,as indicated by the dashed line.}
    \label{fig:benchmark_times}
\end{figure}

Having established the accuracy of the trained NN, the next step was to deploy it in MD simulations. To further assess the viability of the MLIP, we evaluated its computational efficiency by measuring the time required to complete $100$ NPT steps for various system sizes. As shown in Figure~\ref{fig:benchmark_times}, the results exhibit a linear scaling behavior with the system size, $N$, ${\cal O} (N)$, confirming that our MLIP maintains computational efficiency even as the number of atoms increases. These findings underscore the model's suitability for large-scale graphene simulations while preserving \textit{ab initio} accuracy.

\begin{figure}
    \centering
    \includegraphics[width=0.5\textwidth]{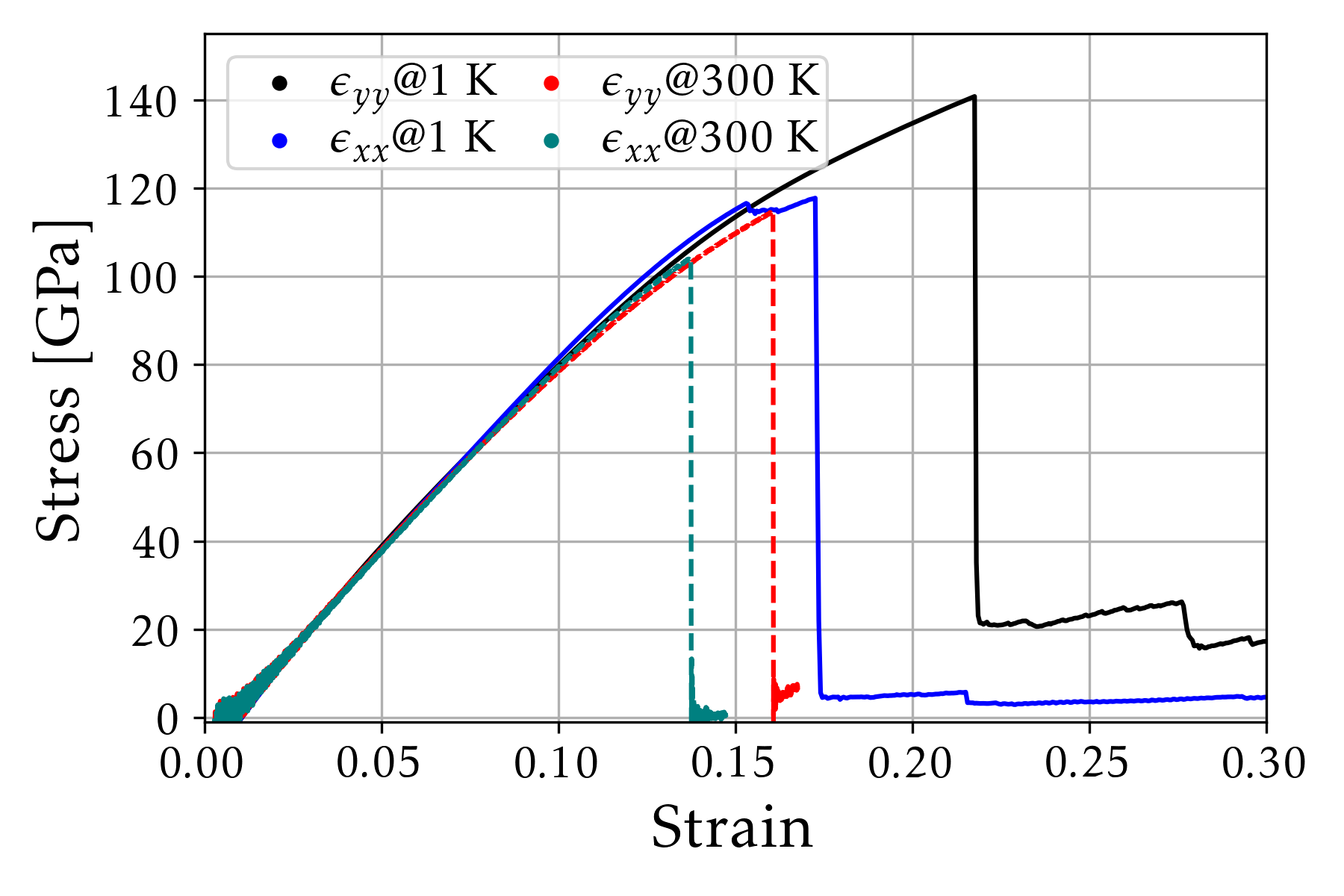}
    \caption{Stress-strain curves obtained using the MLIP. Solid lines correspond to simulations at $T=1~$K, while dashed lines represent results at $T=300~$K.}
\label{fig:strain}
\end{figure}

\textbf{Strain Engineering}. Figure \ref{fig:strain} depicts the stress-strain curves at two different temperatures ($1$ and $300~$K), with stress applied independently along both directions. The MLIP successfully reproduces two expected results related to temperature and strain direction~\cite{Zhao2010}. The temperature effect is evident in the difference in ultimate tensile strength for the same direction and different temperatures, with the fracture occurring earlier at higher temperatures. The directional effect appears in the curves at a given temperature: graphene strained along the $\vec{y}$ direction (i.e., armchair orientation) exhibits greater ductility than when strained along the $\vec{x}$ direction (zigzag orientation).

\begin{figure}
    \centering
   \includegraphics[width=0.5\textwidth]{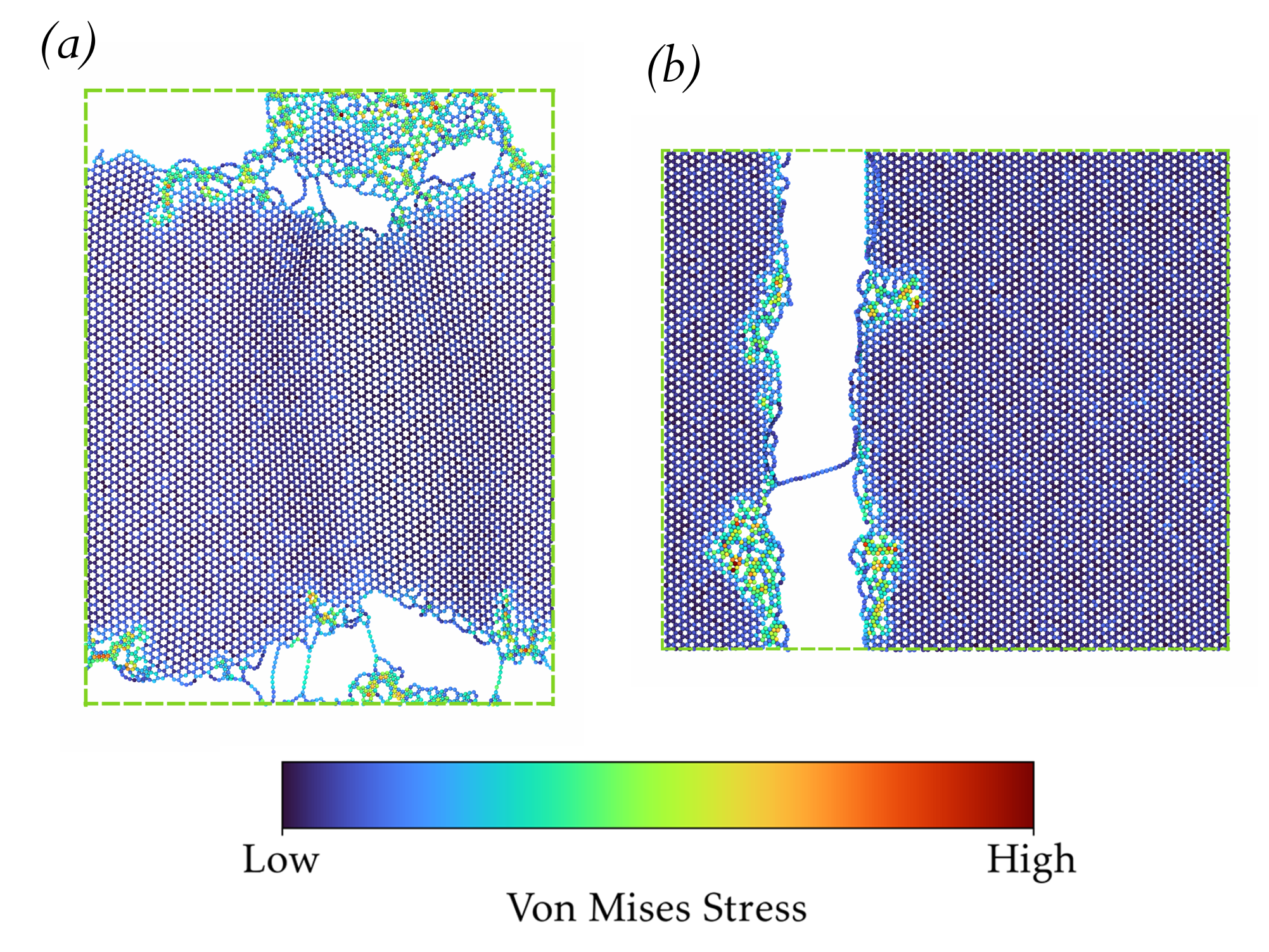}
    \caption{Snapshots from strain simulations for $T=300~$K, with strain $(a)$ $\epsilon_{yy}$ and $(b)$ $\epsilon_{xx}$. Atoms are colored according to their Von Mises stress.}
    \label{fig:snapshots}
\end{figure}

Snapshots of the torn graphene sheets for $T=300~$K are shown in Figure~\ref{fig:snapshots}, where we note, particularly in Figure~\ref{fig:snapshots}b, the formation of linear acetylenic carbon (LAC) chains~\cite{Bryce2021,Zhang2020} after the graphene sheets fractured.

Even though LACs exhibit mechanical properties that differ significantly from those of graphene~\cite{Kotrechko2015}, their formation has been closely associated with applying strain to graphene. Previous studies have reported LAC formation, either by individually pulling atoms~\cite{Wang2007} or by assembling carbon atoms atop one another~\cite{Ataca2011}, suggesting a close connection between mechanical deformation and the emergence of carbyne. We also note that these structures appeared in the torn regions of the graphene sheets at all temperatures, with a more pronounced presence when stress was applied in the $\vec{x}$ direction, as shown in the videos available in the Supporting Information.
\begin{table*}[h]
\centering
\caption{Comparison of elastic properties obtained in this work with values from previous studies. The table includes the elastic constants $C_{11}$ and $C_{21}$, shear modulus $\gamma_{12}$, Poisson’s ratio, $\nu$, and Young’s modulus $Y_{m}$. All dimensional quantities are reported in TPa. MM stands for Molecular Mechanics.}

\begin{tabular*}{\textwidth}{@{\extracolsep{\fill}}cccccc}
\toprule
\toprule
$C_{11}~$  & $C_{21}~$ & ${\gamma_{12}}~$ & $\nu$ & $Y_m~$ & Method\\
\midrule
$1.093$  & $0.199$ & $0.446$ & $0.183$ & $1.056$  & MLIP (this work)\\
$0.977$& $0.255 $  & $0.358$ & $0.26$  & $0.912$  & MD (AMBER)~\cite{TSAI2010194}\\

$-$ & $-$  & $0.384$ & $0.416$ & $0.669$ & MD (Tersoff)~\cite{Reddy2006}\\
$-$       & $-$        & $-$       & $-$       & $0.967$ & MD (AIREBO)~\cite{Liang2022} \\
$-$      & $-$       & $-$       & $0.162$ & $1.096$  & Experiment~\cite{Fan2017}\\
$-$ & $-$ & $-$ & $0.186$ & $1.050$&  DFT~\cite{Liu2007}\\
$-$ & $-$ & $-$ & $0.160$ & $1.060$&  DFT~\cite{Gouadec2007} \\
$0.861$ & $0.342$ & $-$ & $0.398$ & $0.725$& MM~\cite{QIANG2009} \\
$-$ & $-$ & $-$ & $-$ & $0.975$&  MLP~\cite{Singh2023} \\
\bottomrule
\bottomrule
\end{tabular*}
\label{tab:comparison_lit}
\end{table*}

The elastic tensor components and derived mechanical properties obtained using our MLIP, i.e., the Young modulus, Poisson's ratio, shear modulus and the constants $C_{11}$ and $C_{21}$, are presented in Table~\ref{tab:comparison_lit} alongside reference values from the literature based on DFT, MD, and experimental studies. Regarding our results, we note the excellent agreement between the Poisson's ratio, $\nu$, and Young modulus, $Y_m$, obtained experimentally and via DFT $-$ Refs.~\citenum{Fan2017} and \citenum{Liu2007,Gouadec2007}, respectively. We also note that the discrepancies observed in the elastic constants and shear modulus, when compared to previous MD results, can be attributed to the known tendency of conventional MD force fields to model graphene as a significantly ``squishier'' material~\cite{TSAI2010194}.

\textbf{Vibrational Analysis}. Finally, we extended our investigation of the MLIP beyond its original training scope, focused on stress/strain properties, to evaluate its predictive capabilities for vibrational and thermal properties. Specifically, we examined the phonon band structure to characterize thermal behavior and analyzed the VDOS--which is inherently linked to Raman spectroscopy--to gain preliminary insights into graphene's optical properties. This analysis provided an initial assessment of the model’s transferability to vibrational phenomena.

\begin{figure}[h]
    \centering
    \includegraphics[width=0.45\textwidth]{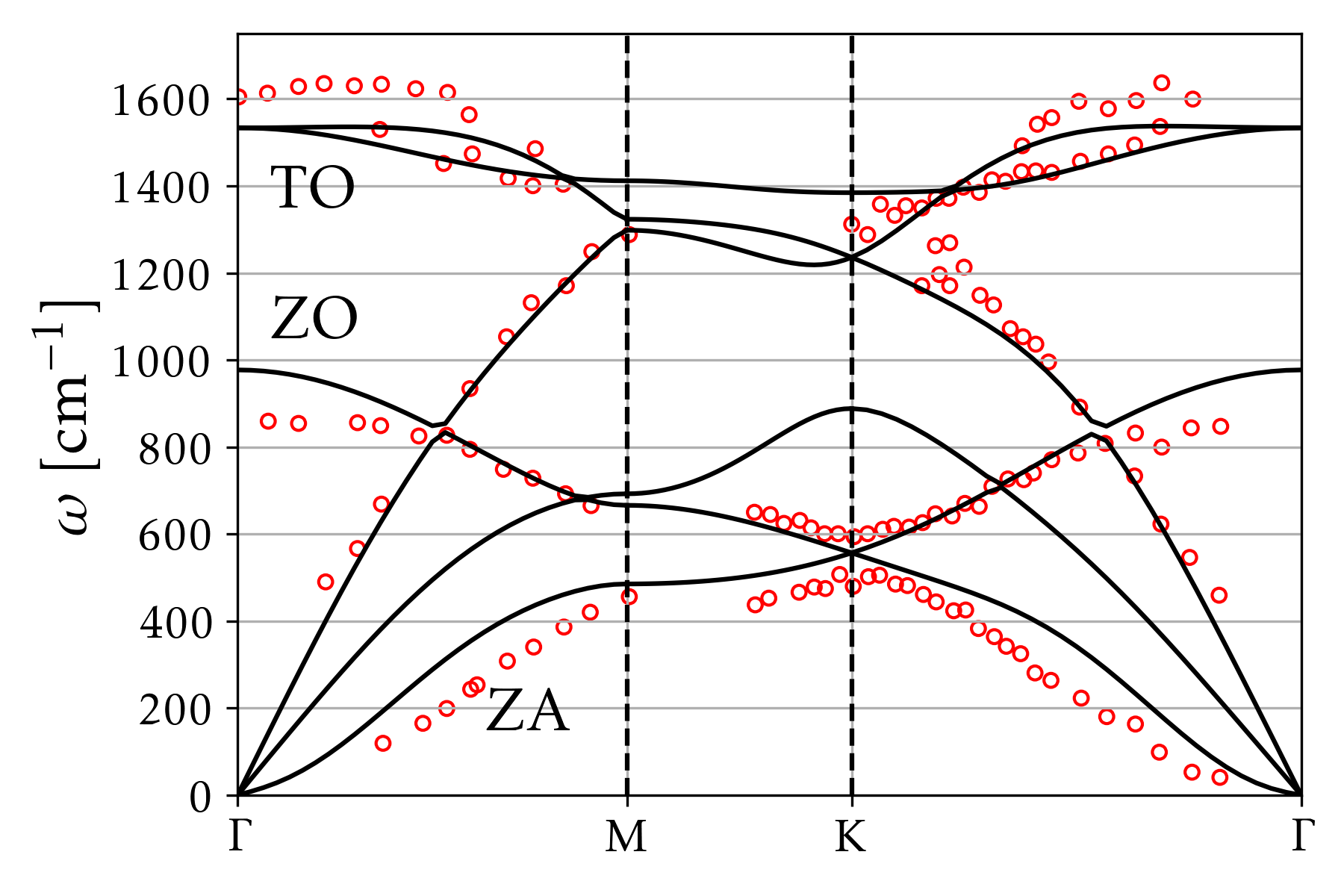}

    \caption{Phonon dispersion of graphene obtained using the MLIP in LAMMPS. The in-plane (TO) and out-of-plane (ZO, ZA) phonon branches are labeled accordingly. Notably, the ZA vibrational mode exhibits quadratic behavior around  $\Gamma$ point. The scattered open circles represent experimental data from Ref.~\cite{Yanagisawa2005}.}
    
    \label{fig:phonons}
\end{figure}

Figure~\ref{fig:phonons} depicts the phonon band structure obtained using the MLIP, as well as experimental points from a high-energy electron energy-loss spectroscopy (HEELS) experiment, reproduced from Yanagisawa et al.~\cite{Yanagisawa2005}. Notably, the MLIP captures the quadratic behavior of the in-plane ZA mode near the $\Gamma$ point, and shows good agreement with the experimental data, specially along the $\text{K}$-$\Gamma$ path.

\begin{figure}[h]
    \centering
    \includegraphics[width=0.45\textwidth]{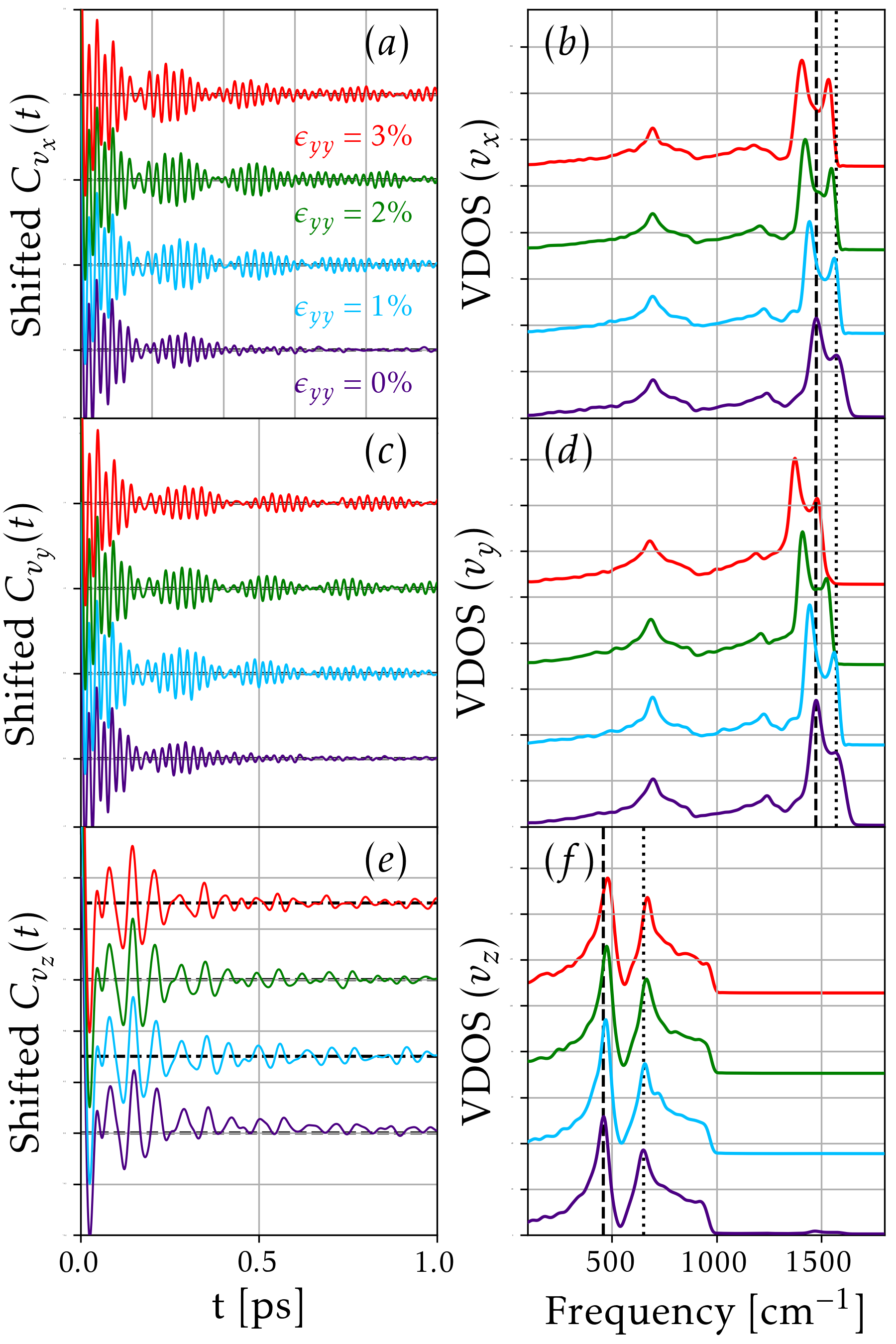}
    \caption{ $C_v(t)$ (left) and corresponding VDOS (right) for $v_x$ (a)-(b), $v_y$ (c)-(d), and $v_z$ (e)-(f). The values in all panels are shifted along the $y$-axis for clarity. The color scheme indicates strain applied along $\vec{y}$, as shown in panel (a).}
    \label{fig:vdos}
\end{figure}

Figure~\ref{fig:vdos} presents the component-separated velocity-velocity autocorrelation function (VACF) and the corresponding VDOS for the $9\times10^3$ carbon atom graphene system, with various strain values along $\vec{y}$. The corresponding results for strain applied in $\vec{x}$ can be found in Figure S2. For the relaxed structures, the two most prominent peaks can be found in Figure~\ref{fig:vdos}b $1570.08\text{cm}^{-1}$ (dotted), $1476.77\text{cm}^{-1}$ (dashed), Figure~\ref{fig:vdos}d $1572.12\text{cm}^{-1}$ (dotted), $1473.75\text{cm}^{-1}$ (dashed) and Figure~\ref{fig:vdos}f $651.81\text{cm}^{-1}$ (dotted), $459.01\text{cm}^{-1}$ (dashed). As indicated by Figure~\ref{fig:vdos}a,c, the in-plane vibrations (along $\vec{x}$ and $\vec{y}$) correspond to higher frequencies, typically associated with the upper region of the phonon dispersion, in agreement with Figure~\ref{fig:phonons}.

We also observed red (blue) frequency shift in Figure~\ref{fig:vdos} for the in-plane (out-of-plane) VDOS upon the introduction of strain. The amplitudes of the shifts, from $\epsilon_{yy} =0 \%$ to $\epsilon_{yy} = 3\%$ on the greatest peaks are $100.06\text{cm}^{-1}$ and $69.75\text{cm}^{-1}$ for  Figure~\ref{fig:vdos}b,d, corresponding to the red-shift and $17.08\text{cm}^{-1}$ for Figure~\ref{fig:vdos}f, corresponding to the blue shift.

Remarkably, the in-plane vibrations, shown in Fig.~\ref{fig:vdos}(a)-(d), align well with previous investigations on graphene's thermal conduction using Nonequilibrium MD~\cite{Liu2014}, as well as \textit{ab initio} calculations on the vibrational spectrum~\cite{Liu2007}. Additionally, our results reproduce the red-shift observed on the (Raman equivalent) G peaks~\cite{Wu2018}, around $\sim1500\text{cm}^{-1}$, under uniaxial strain, consistent with experimental observations~\cite{Zheng2015,Dong2017}.

Overall, the strong agreement between the shifts observed here and other works in the literature validates the MLIP’s ability to capture vibrational anharmonicity and strain-induced effects, key features for reliable predictions of thermal conductivity and phonon-limited transport phenomena.

\section{Conclusions}
\label{sec:conclusion}

In this work, we developed a MLIP for graphene based on the Deep Potential framework. The model was trained using a diverse and comprehensive dataset generated via AIMD simulations under various thermodynamic conditions and strain regimes. This enabled the construction of a reactive, accurate, and computationally efficient potential that captures key mechanical and vibrational properties.

The MLIP demonstrated excellent agreement with AIMD and experimental data for stress–strain responses, elastic constants, phonon dispersion, and VDOS. Notably, it accurately captured temperature-dependent fracture behavior and the formation of linear acetylenic carbon (LAC) chains during tearing—an emergent phenomenon not trivially reproduced by traditional force fields. Additionally, the model correctly predicted key phonon features, such as the quadratic ZA mode and strain-induced red and blue shifts in the VDOS, highlighting its robustness and transferability.

Thanks to its linear computational scaling and high fidelity, the proposed MLIP enables large-scale molecular dynamics simulations of graphene with \textit{ab initio}-level accuracy, making it a valuable tool for investigating thermal, mechanical, and phonon-mediated transport phenomena.

All training data and model parameters are publicly available through the NOMAD repository~\cite{hawthorne2025nomad}, and the trained MLIP is hosted on GitHub to facilitate reproducibility and further developments~\cite{hawthorne2025graphenemlip}.

Future extensions may include expanding the training dataset to cover a wider range of temperature and defect configurations, as well as applying the workflow to other two-dimensional materials, paving the way for accurate, large-scale ML-driven simulations across diverse nanomaterials.

\begin{acknowledgement}
The authors would like to thank CNPq and CAPES for financial support, the Center for Computing in Engineering and Sciences at Unicamap and the HPC Cluster Coaraci, made available under FAPESP grants 2013/08293-7 and 2019/17874-0, respectively, the Brazilian LNCC (Laboratório Nacional de Computação Científica), for access to the SDumont Cluster, under the SINAPAD/2014 project 01.14.192.00. The authors also gratefully acknowledge the computing time made available to them on the high-performance computer ``Lise'' at the NHR Center NHR@ZIB. This center is jointly supported by the Federal Ministry of Education and Research and the state governments participating in the NHR (\url{www.nhr-verein.de/unsere-partner}).

\end{acknowledgement}



\bibliography{refs}






\end{document}


\newpage

\subsection{Visualization of the structures}

\begin{figure*}
    \centering
    \includegraphics[width=\textwidth]{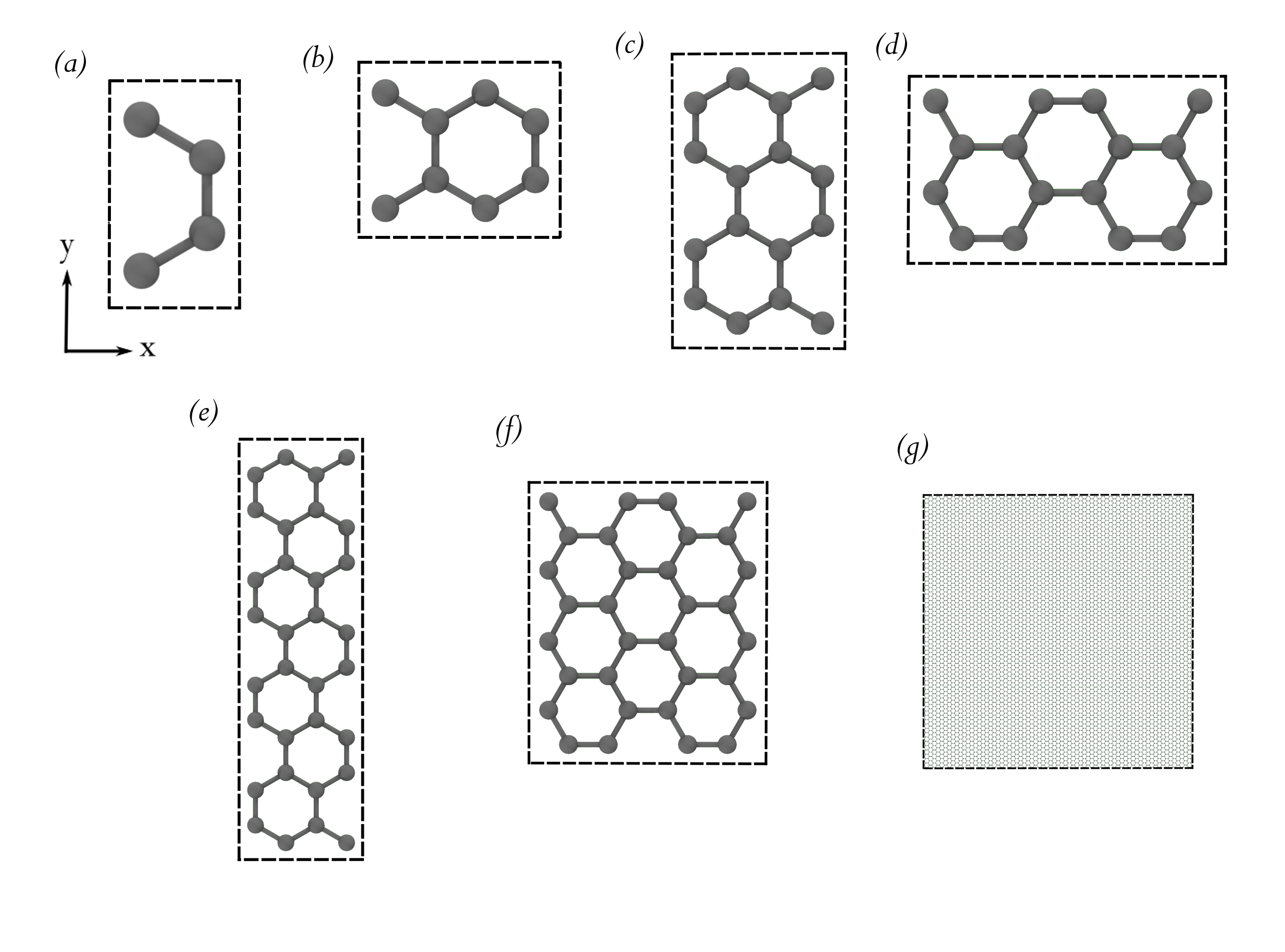}
    \caption{Illustrations of the graphene structures used in the AIMD simulations, containing $(a)~4,~(b)~8,~(c)~16,~(d)~16,~(e)~32$, $(f)~32$ and $(g)~9072$ carbon atoms. The axis illustrated in panel $(a)$ is shared across all others.}
    \label{fig:all_geoms}
\end{figure*}

\newpage

\subsection{Detailed Dataset Frames}

\begin{table*}[h]
\centering
\caption{Detailed information about the Born-Oppenheimer Molecular Dynamics performed to generate the dataset used to train the MLIP. The notation $\epsilon_{\alpha\beta}$ denotes the strain rate applied in the $\alpha\beta$ direction. The \#disp./time values indicate how many different geometries were used in the simulations, and for how much time each of these simulations were performed on average.}
\label{table:datasets}
\hspace{-1.7cm}
\begin{tabular*}{1.1\textwidth}{|c|c|c|c|c|c|}

\cline{1-6}
$\#$-C & \multicolumn{3}{c|}{NPT ($T=300$K)} & NVT ($T=300$K) & NVE \\
\cline{1-6}
 &  $\epsilon_{xx}$/time & $\epsilon_{yy}$/time  & \# disp./time& \# disp./time & \# disp./time \\
\cline{1-6}
4 & $10^{-6}~\text{fs}^{-1}$/$2~\text{ps}$  & $10^{-6}~\text{fs}^{-1}$/$2~\text{ps}$  &  &  & \\
& $10^{-5}~\text{fs}^{-1}$/$2~\text{ps}$  & $10^{-5} \text{fs}^{-1}$/$2~\text{ps}$  & $10$/$2~\text{ps}$ &$10/2~\text{ps}$& $10/2.2~\text{ps}$\\
& $10^{-4}~\text{fs}^{-1}$/$2~\text{ps}$  & $10^{-4} \text{fs}^{-1}$/$2~\text{ps}$  &  & & \\
& $10^{-3}~\text{fs}^{-1}$/$2~\text{ps}$  & $10^{-3} \text{fs}^{-1}$/$2~\text{ps}$  &  & & \\
\cline{1-6}
8 & $10^{-6} \text{fs}^{-1}$/$3~\text{ps}$  & $10^{-6} \text{fs}^{-1}$/$3~\text{ps}$  &  &    & \\
& $10^{-5} \text{fs}^{-1}$/$3~\text{ps}$  & $10^{-5} \text{fs}^{-1}$/$3~\text{ps}$  & 10/$3~\text{ps}$ &$10/4.1~\text{ps}$& $10/2.2~\text{ps}$\\
& $10^{-4} \text{fs}^{-1}$/$3~\text{ps}$  & $10^{-4} \text{fs}^{-1}$/$3~\text{ps}$  &  & & \\
& $10^{-3} \text{fs}^{-1}$/$3~\text{ps}$  & $10^{-3} \text{fs}^{-1}$/$3~\text{ps}$  &  & & \\
\cline{1-6}
16 & $10^{-6} \text{fs}^{-1}$/$1.8~\text{ps}$  & $10^{-6} \text{fs}^{-1}$/$1.8~\text{ps}$  &  &  & \\
$~(L_{y>})$& $10^{-5} \text{fs}^{-1}$/$1.8~\text{ps}$  & $10^{-5} \text{fs}^{-1}$/$1.8~\text{ps}$  & 10/$2~\text{ps}$ &$10/1.6~\text{ps}$& $10/2.2~\text{ps}$\\
& $10^{-4} \text{fs}^{-1}$/$1.8~\text{ps}$  & $10^{-4} \text{fs}^{-1}$/$1.8~\text{ps}$  &  & & \\
& $10^{-3} \text{fs}^{-1}$/$1.8~\text{ps}$  & $10^{-3} \text{fs}^{-1}$/$1.8~\text{ps}$  &  & & \\
\cline{1-6}
16 & $10^{-6} \text{fs}^{-1}$/$0.6~\text{ps}$  & $10^{-6} \text{fs}^{-1}$/$0.6~\text{ps}$  &  &  &\\
$~(L_{x>})$& $10^{-5} \text{fs}^{-1}$/$0.6~\text{ps}$  & $10^{-5} \text{fs}^{-1}$/$0.6~\text{ps}$  & 10/$1~\text{ps}$ &$10/0.5~\text{ps}$&  $10/1.1~\text{ps}$\\
& $10^{-4} \text{fs}^{-1}$/$0.6~\text{ps}$  & $10^{-4} \text{fs}^{-1}$/$0.6~\text{ps}$  &  & & \\
& $10^{-3} \text{fs}^{-1}$/$0.6~\text{ps}$  & $10^{-3} \text{fs}^{-1}$/$0.6~\text{ps}$  &  & & \\
\cline{1-6}
32 &   & $10^{-6} \text{fs}^{-1}$/$1.5~\text{ps}$  & & & \\

$~(L_{y>})$&  &   & && \\

& $10^{-4} \text{fs}^{-1}$/$5~\text{ps}$  & $10^{-4} \text{fs}^{-1}$/$3.8~\text{ps}$  &  & $1/1~\text{ps}$&$1/1~\text{ps}$ \\
\cline{4-4}
&   & $5\times10^{-3} \text{fs}^{-1}$/$0.5~\text{ps}$  & $\epsilon_{xy}$/time  & & \\
& $10^{-2} \text{fs}^{-1}$/$0.5~\text{ps}$  &  &  $10^{-4}\text{fs}^{-1}$/0.5~\text{ps} & & \\
\cline{1-6}

32 &   & $10^{-6} \text{fs}^{-1}$/$1.5~\text{ps}$  & & & \\

$~(L_{x>})$&  &   & && \\

& $10^{-4} \text{fs}^{-1}$/$5~\text{ps}$  & $10^{-4} \text{fs}^{-1}$/$3.8~\text{ps}$  &  & $1/1~\text{ps}$&$1/1~\text{ps}$ \\
\cline{4-4}
&   & $5\times10^{-3} \text{fs}^{-1}$/$0.5~\text{ps}$  & $\epsilon_{xy}$/time  & & \\
& $10^{-2} \text{fs}^{-1}$/$0.5~\text{ps}$  &  &  $10^{-4}\text{fs}^{-1}$/0.5~\text{ps} & & \\
\cline{1-6}
\end{tabular*}
\end{table*}

\newpage

\subsection{$C_v(t)$ and VDOS for $\epsilon_{xx}$}

\begin{figure*}[h]
    \centering
    \includegraphics[width=0.8\textwidth]{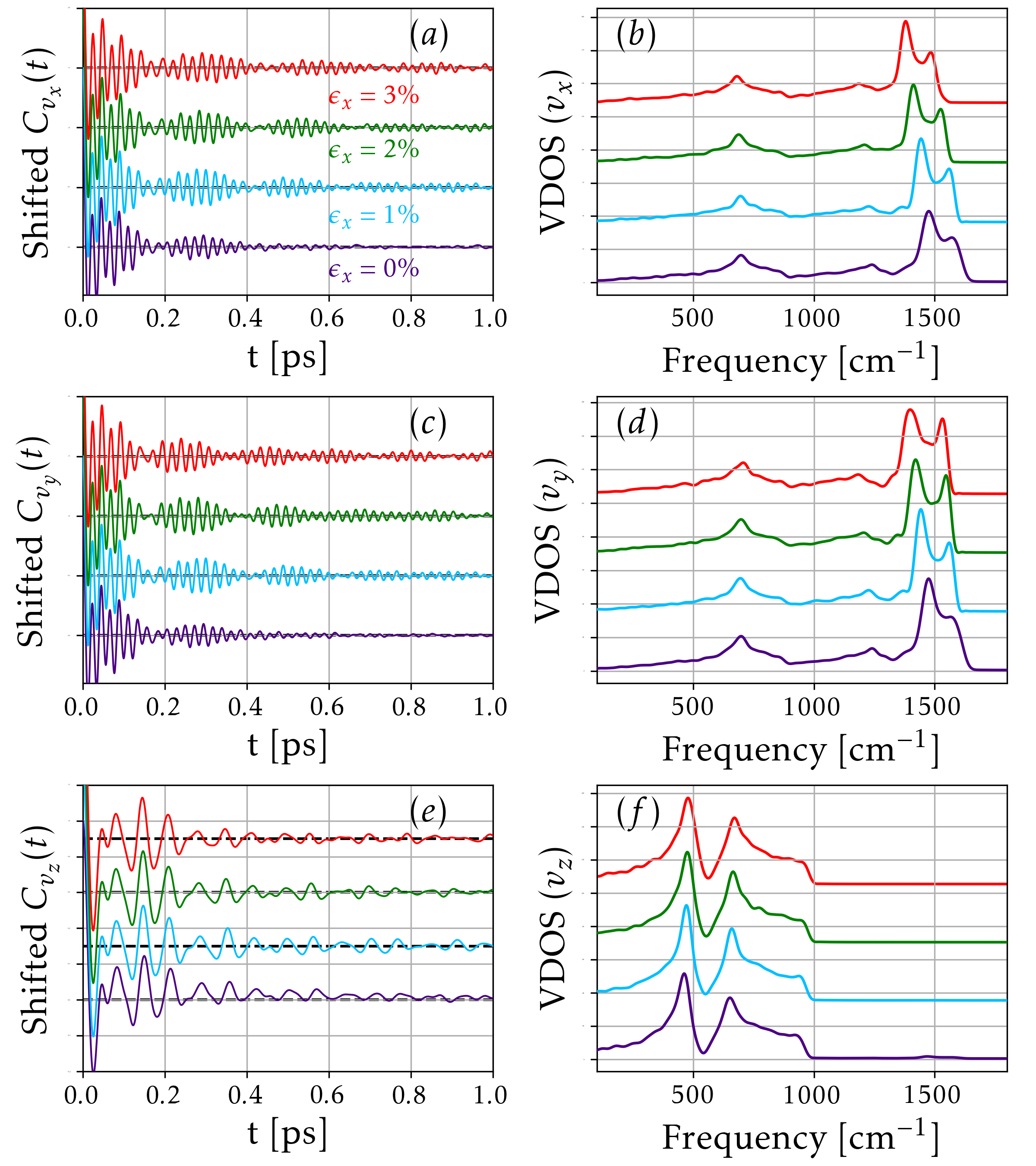}
    \caption{Corresponding velocity autocorrelation functions and VDOS (left and right panels, respectively) for graphene strain with $\epsilon_{xx}$.}
    \label{fig:vdos-si}
\end{figure*}

\subsection{Link to the videos}

The videos of the strain applications at $T=1~$K and $T=300~$K can be found here: \url{https://www.youtube.com/watch?v=dauxu3UeUh4}.